\documentclass[aps,prl,twocolumn,amsmath,superscriptaddress]{revtex4-2}

\usepackage[urlcolor=blue,colorlinks=true,citecolor=blue,linkcolor=blue,pdfstartview={FitH},bookmarks=false]{hyperref}
\usepackage{graphicx}
\usepackage{longtable}
\usepackage{epsfig}
\usepackage{dcolumn}
\usepackage{bm}
\usepackage{amssymb}
\usepackage{multirow}
\usepackage{times,xcolor}
\usepackage{hyperref}
\usepackage{braket}
\usepackage{comment}
\usepackage{subfigure}
\usepackage{MnSymbol}

\newcommand{\cdag}{c^{\dagger}}
\newcommand{\cnod}{c^{\phantom{\dagger}}}
\newcommand{\adag}{a^{\dagger}}
\newcommand{\anod}{a^{\phantom{\dagger}}}
\newcommand{\hdag}{h^{\dagger}}
\newcommand{\hnod}{h^{\phantom{\dagger}}}
\newcommand{\bdag}{b^{\dagger}}
\newcommand{\bnod}{b^{\phantom{\dagger}}}
\newcommand{\atanh}{\textrm{atanh}}

\newcommand{\bvec}[1]{\mathbf{#1}}

\begin{document}
\title{Altermagnetic polarons: the fate of alter magnetic band splittings at strong coupling}

\author{Maria Daghofer}
\email{maria.daghofer@fmq.uni-stuttgart.de}
\affiliation{
Institut f\"ur Funktionelle Materie und Quantentechnologien,
Universit\"at Stuttgart,
70550 Stuttgart,
Germany}

\author{Krzysztof Wohlfeld}%
\affiliation{Institute of Theoretical Physics, Faculty of Physics, University of Warsaw, Pasteura 5, PL-02093 Warsaw, Poland}

\author{Jeroen van den Brink}
\affiliation{Leibniz-Institut f\"{u}r Festk\"{o}rper- und Werkstoffforschung, Helmholtzstraße 20, D-01069 Dresden, Germany}
\affiliation{Institute of Theoretical Physics and W{\"u}rzburg-Dresden
  Cluster of Excellence {\it ct.qmat}, Technische Universit{\"a}t Dresden,
  01062 Dresden, Germany}

\begin{abstract}
While a spin-dependent band splitting is one of the
characteristic features of altermagnets, the conventional band picture
itself breaks down in the many altermagnets that are correlated Mott
materials. We employ two numerical many-body methods, the
self-consistent Born approximation and variational cluster approach,
to explore this strongly correlated regime and investigate hole motion
in Mott altermagnets. 
Our results reveal that spin-dependent spectral-weight transfer is the
dominant signature of Mott altermagnetism. This  pronounced
spin–momentum locking of the quasiparticle spectral  weight arises
from the formation of altermagnetic polarons, whose  dynamics are
governed by the interplay between free hole motion and  the coupling
of the hole to magnon excitations in the  altermagnet. We demonstrate
this effect by calculating ARPES spectra  for 
three canonical
altermagnetic systems: the checkerboard  $J$–$J'$ model, a variant describing the transition-metal--ion sites of the inverse Lieb lattice, and the
Kugel–Khomskii spin–orbital altermagnet based on cubic vanadates
$R$VO$_3$ ($R$=La, Pr, Nd,  Y).  
\end{abstract}
\date{\today}

\maketitle

It was recently discovered that, depending on their symmetries,
collinear antiferromagnets may break spin degeneracy in momentum
space, even in the absence of spin-orbit coupling, giving rise to a
set of magnetic materials referred to as altermagnets
(AMs)~\cite{Smejkal20,Smejkal22a,Naka19,Yuan20,Naka21,Guo23}. At a
single-particle level, valid for systems with weak electron-electron
interactions, this gives rise to electronic bands with anisotropic,
spin-split iso-energy surfaces. The anisotropic bands in turn lead to
a series of unusual electric, magneto-electric and optical
properties~\cite{Smejkal22,Sato24,McClarty24,2025arXiv250220010D}.  
Indeed, such spin-split bands corresponding to 
(very)
weak
correlations without band renormalization have been observed in
angle-resolved photoemission (ARPES) experiments, for instance in metallic CrSb or semiconducting
MnTe~\cite{Reimers24,Yang25,Zeng24,Ding24,Li24,Krempasky24,Osumi24}. 

However, a large set of AMs are actually strongly correlated Mott
insulators, with band gaps determined by the Hubbard electron-electron
interaction~\cite{Guo23}. This comprises, e.g., a series of transition metal fluoride AMs with
rutile structure~\cite{Guo23}, vanadates~\cite{CUONO2023171163}, 
recently discovered layered AM La$_2$O$_3$Mn$_2$Se$_2$
\cite{Wei24,2025arXiv250621661G}, and potentially 
transition metal 
  oxychalcogenides~\cite{PhysRevLett.104.216405,s31h-hk2v}. It is well known that in
canonical correlated 
antiferromagnets the band-picture fails as 
many-body effects strongly renormalize quasiparticle
dispersions all the way down to the magnetic energy scale, and at the
same time vehemently reorganize spectral weights (SWs)~\cite{Damascelli2003, Lee2006}. This raises the
question as to the fate of altermagnets and their dynamical properties
in the strongly correlated and Mott insulating regime. 

\begin{figure}
    \subfigure[]{\includegraphics[width=0.49\columnwidth, trim= 0 0 0 0 , clip]{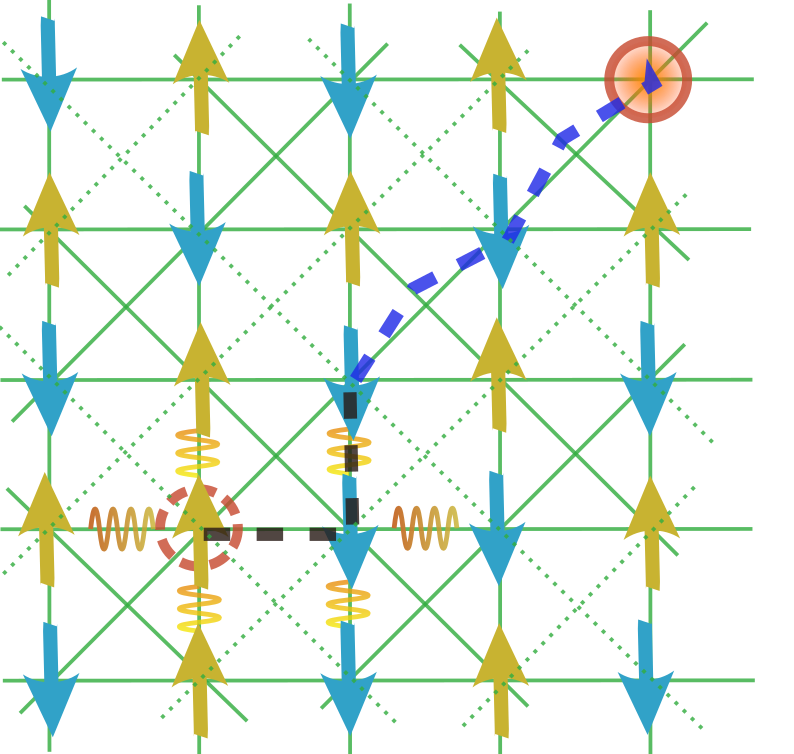}\label{fig:cartoon_checker}}
    \subfigure[]{\includegraphics[width=0.49\columnwidth, trim= 0 40 160 0 , clip]{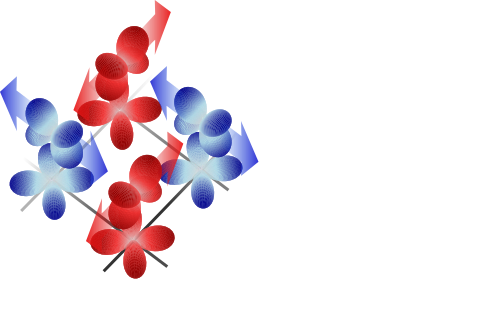}\label{fig:cartoon_LaVO3}}
  \caption{(a) Cartoon for hole propagation in the (generalized) checkerboard model: The hole has moved from its
    original position (dashed empty circle) to the current one (filled circle)
    via two NN hoppings (black dashed horizontal and vertical line) and two
    NNN hopping processes (blue dashed diagonal lines). The NN processes have
    disturbed the AF background and thus created magnons, the NNN hopping left
    it intact. Adding (different) couplings along the dotted
      lines as well descibes the transition-metal sublattice of the inverse
      Lieb lattice (ILL)~\cite{2025arXiv250804839C}. (b) Cartoon for the spin-orbital model based on
    LaVO$_3$. The 
    $xy$ orbital is half filled and superexchange leads to AF
    order, indicated by red and blue color. Additionally, the $xz$/$yz$ orbitals share one electron, which can
    only hop along one direction (indicated by the fat arrows) and whose spin is aligned to the $xy$ spin on
    the same site. Superexchange processes due to $xz$/$yz$ hopping induce
    Ising spin coupling.
  \label{fig:cartoon}}
\end{figure}

In strong-coupling correlated 
antiferromagnets the large difference with the bare
single-particle electronic dispersions is due to the disruptions in
the antiferromagnetic (AF) background that are caused by a  moving
charge carrier. Its motion dynamically creates and annihilates
excitations of the AF background, usually magnons, and the interplay
of hole and magnons strongly modifies hole propagation. While a
quasi-particle (QP) -- the spin polaron -- often survives, it contains
only part of the SW and has a dispersion of the order of
the magnetic exchange $J$ rather than of the bare hopping
$t$~\cite{PhysRevB.39.6880}. The missing weight is lost into an
incoherent continuum~\cite{PhysRevLett.60.2793, Martinez1991} that is usually
separated from the polaronic QP band and distributed over a width
$\propto t$. The description of charge carriers in terms of spin
polarons in antiferrromagnets is widely used~\cite{Wang2015, Wang2020,
  Bacq-Labreuil2025} to interpret ARPES measurements on
cuprates~\cite{Ronning2005, Wells1995} and has been  
extended to describe 
orbital excitations in iridates~\cite{ir_QP} and
cuprates~\cite{Martinelli2024}, as well as been investigated in
ultracold-atom quantum
simulators~\cite{PhysRevX.11.021022,cold_atom_pol, Miller2025}.  

Certain theoretical aspects of AMs beyond the weak-coupling regime
have been considered, in particular the existence of lower and upper
Hubbard bands in addition to altermagnetic band
splitting~\cite{PhysRevB.111.L020401}, the persistence of
altermagnetic signatures into a frustrated and/or fluctuating
regime~\cite{PhysRevB.110.205140,PhysRevResearch.7.023152},
  and non-trivial temperature effects~\cite{7nvm-s225}. However, the  
impact of altermagnetic symmetries on the properties of strongly interacting
quasi-particles and spin polarons, which are directly related to the
spectral function measured in ARPES, have not been considered so far. 

Here we consider the correlated charge propagation in Mott-insulting
altermagnets, i.e. the spin polaron in an altermagnet instead of an
antiferromagnet. Both the bare dispersion, which one would observe without
correlations, and the magnon
spectrum~\cite{PhysRevLett.131.256703,Liu2025}, which strongly
modifies the effective 
dispersion of a spin polaron, have momentum-dependent band splittings in
altermagnets that are not present in antiferromagnets. Focusing on a $t$-$J$
model on the checkerboard lattice, we find that momentum and spin of the
polaron are similarly strongly linked. However, in contrast to weakly
interacting altermagnets, this is not necessarily reflected in a band
splitting, but rather in the particle life time: One spin projection is pushed
into the incoherent spectrum, loosing its quasi-particle character, so that
the surviving quasi-particle shows spin-momentum locking. 

We also look at hole motion in a Mott-insulating Kugel-Khomskii-type model for vanadates of the form RVO$_3$ (R=La, Pr, Nd, Y)  
that combine orbital and magnetic order~\cite{Fujioka2005,Miyasaka2000,
  PhysRevB.106.115110}. Orbital order is driven by a combination
of lattice and superexchange effects, with LaVO$_{3}$ being most clearly dominated by
superexchange~\cite{PhysRevB.106.115110}. Looking at a single two-dimensional
plane, spin and orbital alternate together in an
antiferromagnetic/antiferroorbital (AF/AO) pattern as sketched in
Fig.~\ref{fig:cartoon_LaVO3}, suggesting an AM state~\cite{PhysRevLett.132.236701}.
In fact, band-structure calculations have found YVO$_{3}$ to be an
altermagnet~\cite{CUONO2023171163},
and we will show that holes are then expected to move as altermagnetic polarons.

\textit{Spectral function of the checkerboard altermagnetic polaron ---}
Our considerations start from a $t$-$J$ model 
on the checkerboard lattice [see Fig.\ref{fig:cartoon_checker}] 
defined by
\begin{align}\label{eq:ham_checker_t}
  H &= -t\hspace{-0.5em}\sum_{\langle i,j\rangle,\sigma}\hspace{-0.5em}
  \cdag_{i,\sigma}\cnod_{j,\sigma}  -t'\hspace{-1em}\sum_{\substack{\llangle i,j\rrangle\parallel (1,1)\\i,j\in A, \sigma}}\hspace{-1em}
  \cdag_{i,\sigma}\cnod_{j,\sigma}
  -t'\hspace{-1em}\sum_{\substack{\llangle i,j\rrangle\parallel (1,-1)\\i,j\in B, \sigma}}\hspace{-1em}
  \cdag_{i,\sigma}\cnod_{j,\sigma}\\
  &\quad +J \sum_{\langle i,j\rangle}
    \bvec{S}_{i}\bvec{S}_{j}
    +J'\hspace{-1em}\sum_{\substack{\llangle i,j\rrangle\parallel (1,1)\\i,j\in A}}\hspace{-1em}
  \bvec{S}_{i}\bvec{S}_{j}
  +J'\hspace{-1em}\sum_{\substack{\llangle i,j\rrangle\parallel (1,-1)\\i,j\in B}}\hspace{-1em}
  \bvec{S}_{i}\bvec{S}_{j}\label{eq:ham_checker_J}\;,
\end{align}
where $\cdag_{i,\sigma}$ ($\cnod_{i,\sigma}$) creates (annihilates) an electron
with spin $\sigma=\uparrow,\downarrow$ on site $i$. $\bvec{S}_{i}$ denotes the
corresponding vector of spin operators
$S^{\alpha}_{i}=\tfrac{1}{2}\sum_{s,s'}\cdag_{i,s}\sigma_{s,s'}^{\alpha} \cnod_{i,s'}$
with Pauli matrices $\sigma^{\alpha}$. Nearest-neighbor (NN) bonds $\langle i,j\rangle$
connect the two sublattices $A$ and $B$. Bonds $\llangle i,j\rrangle$ connect
sites from the same sublattice and are active only along direction $(1,1)$ for
the $A$ sublattice and $(1,-1)$ for $B$, see the solid lines in Fig.~\ref{fig:cartoon_checker}.
NN hopping $t$ is used as unit of energy. 

We employ the self-consistent Born approximation
(SCBA)~\cite{PhysRevB.39.6880} by largely following
Refs.~\cite{PhysRevB.44.317,PhysRevB.78.214423,PhysRevB.79.224433}.
Throughout this work, we remain in the regime of dominant NN exchange, so
  stabilizing N\'eel magnetic order~\cite{Liu_2019}. The
spin-part (\ref{eq:ham_checker_J}) of the  Hamiltonian becomes 
\begin{align}\label{eq:H_J_eff}
  \tilde{H}_{J}=
  \omega_{\alpha,\bvec{k}} \alpha_{\bvec{k}}^\dagger\alpha_{\bvec{k}}^{\phantom{\dagger}}
  +\omega_{\beta,\bvec{k}} \beta_{\bvec{k}}^\dagger\beta_{\bvec{k}}^{\phantom{\dagger}}\;,
\end{align}
where $\alpha_{\bvec{k}}^\dagger$ ($\alpha_{\bvec{k}}^{\phantom{\dagger}}$)
and $\beta_{\bvec{k}}^\dagger$ ($\beta_{\bvec{k}}^{\phantom{\dagger}}$) create
(annihilate) magnons of the two branches, see~\cite{SM} 
for more details as well as
the derived expression for the kinetic part~\eqref{eq:ham_checker_t}.

The primary advantage of SCBA is that it very accurately captures the
unavoidable coupling between the hole and the collective excitations
from the ordered state---the magnons. More precisely, the  
NN hole hopping between
sublattices always creates or annihilates magnons and thus couples hole motion and
magnons, see~Fig.~\ref{fig:cartoon_checker}. 
It is then only the subdominant
NNN hole hopping within a
sublattice that does not disturb the AM background,
see~Fig.~\ref{fig:cartoon_checker}. The SCBA captures both of these processes
in the self-consistent equations
for the self energy 
\begin{align}\label{eq:sigma_SCBA}  
  \Sigma^{\alpha}(\bvec{k},\omega) =\frac{z^{2}t^{2}}{N_{u}}
  \sum_{{\bvec{k}}'}M_{\bvec{k},\bvec{k}'}^{2}
  G^{\bar{\alpha}}(\bvec{k}-\bvec{k'},\omega-\omega_{\bar{\alpha},\bvec{k}'})\;,
\end{align}
for sublattice $\alpha = A,B$. As NN hopping changes sublattice, the
Green's function $G^{\bar{\alpha}}$ entering the sum is the one of the
opposite sublattice $\bar{\alpha}=B,A$ and the energy difference
$\omega_{\bar{\alpha},\bvec{k}'}$ is the 
energy of the emitted and re-absorbed $\alpha$- ($\beta$-) magnon for
$\Sigma^{B}$ ($\Sigma^{A}$). The vertex $M_{\bvec{k},\bvec{k}'}$ is given in
the SM~\cite{SM}, $z=4$ is the coordination number and $N_{u}$ the
number of unit cells. The Green's function of obtained from the self
energy via the Dyson equation
\begin{align}
  \left[G^{\alpha}(\bvec{k},\omega)\right]^{-1} = {\omega - \epsilon^{\alpha}(\bvec{k}) - \Sigma^{\alpha}(\bvec{k},\omega)+i\delta}\;,
q\end{align}
with the free dispersion $\epsilon^{\alpha}(\bvec{k}) = 2t'\cos
k_{\alpha} $ coming from NNN hopping. Higher-order processes are here
effectively included, as long as they do not contain crossing magnon
lines. For square-lattice AF states, the lowest crossing terms
vanish, and the impact of the non-crossing approximation has been found
to be negligible~\cite{PhysRevB.45.2425}.
Finally, we again include quantum fluctuations when going from the
  sublattice-dependent Green's function $G^{\alpha}$ to the spin
  dependent $G^{\sigma=\uparrow,\downarrow}$~\cite{SM}.

\begin{figure*}
   \includegraphics[height=0.42\columnwidth,trim= 0 0 80 0, clip]{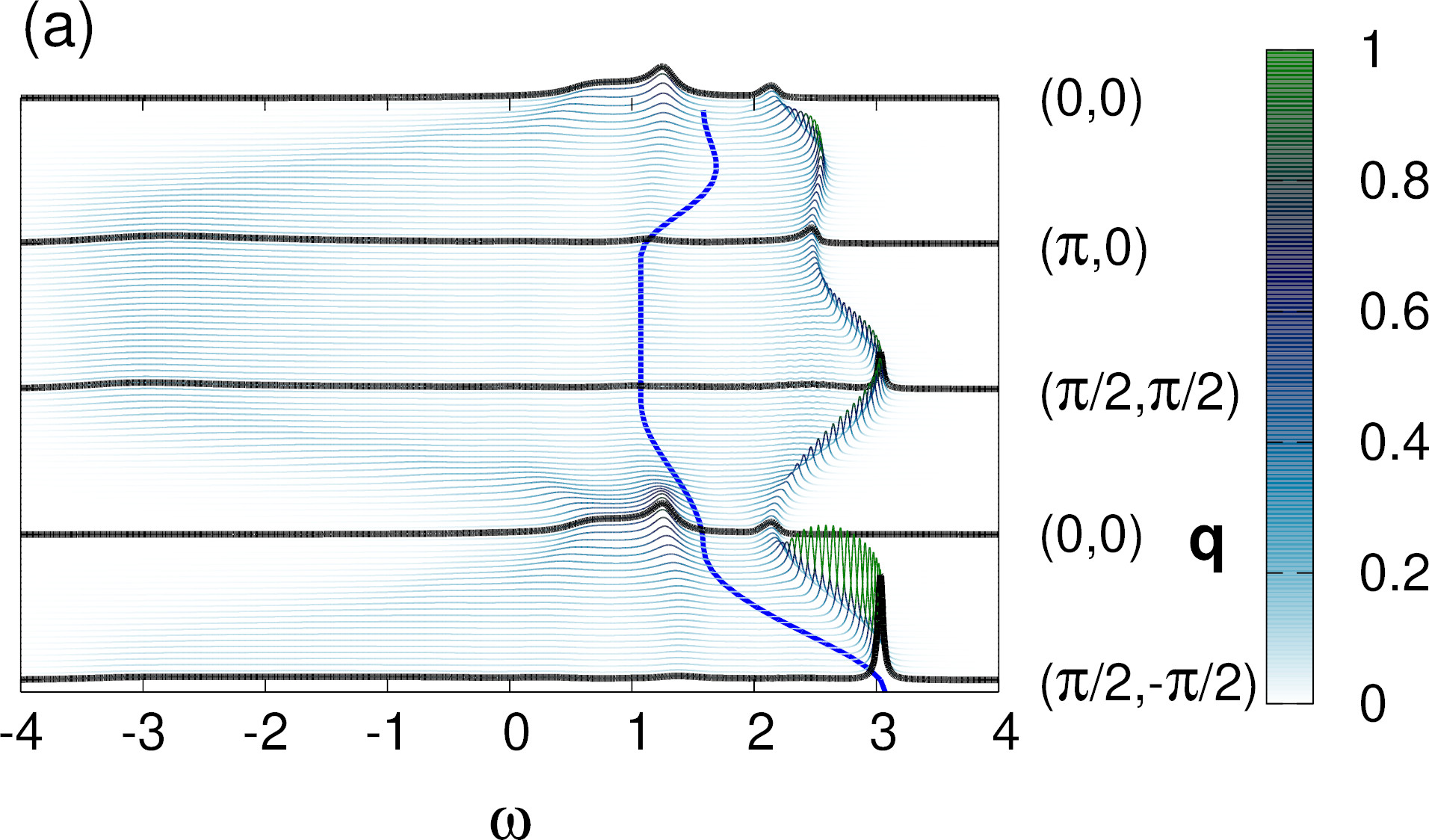}
   \qquad 
   \includegraphics[height=0.42\columnwidth,trim= 0 0 0 0, clip]{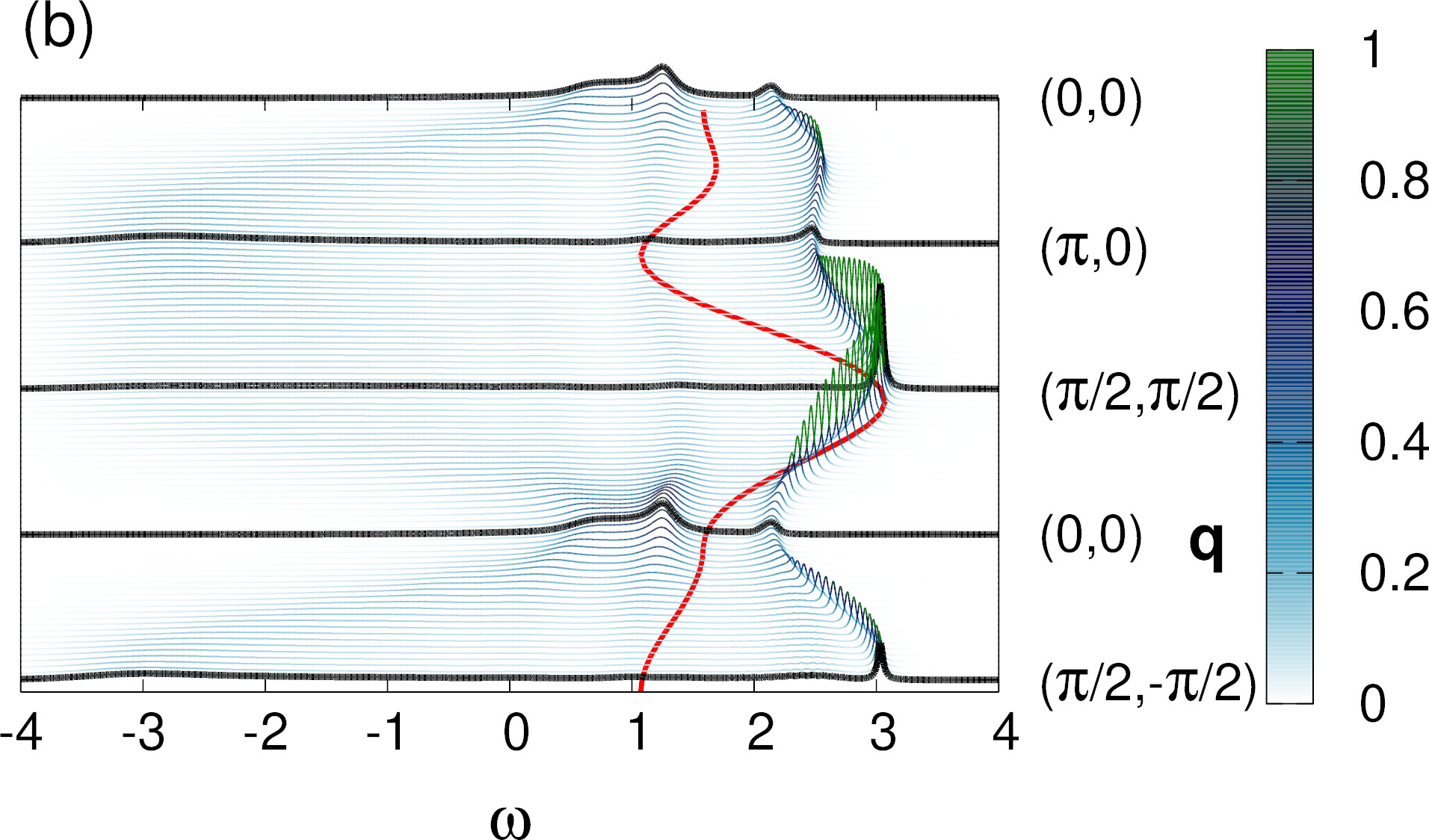}
   \quad
   \includegraphics[height=0.42\columnwidth,trim= 0 0 0 0, clip]{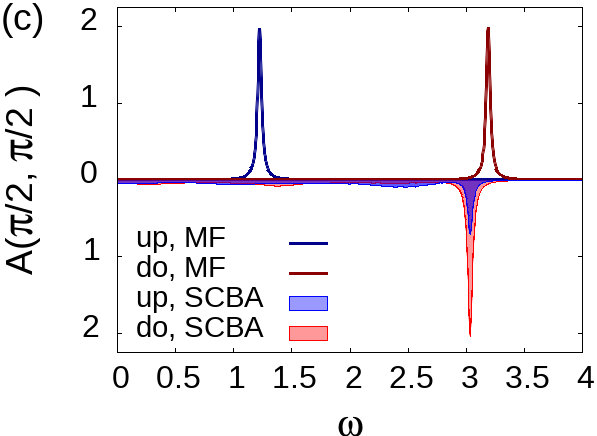}\\
 \caption{One-particle spectral density obtained with SCBA for the checkerboard model with $J=0.4\;t$, NNN $J'=0.15\;t$ and $t'=-0.5\;t$ and (a) spin up, (b) spin down. Momenta are here given in terms of a one-site unit cell; peaks were broadened with a Lorentzian of width $0.02\;t$ for plotting. 
 For comparison to the usual square-lattice antiferromagnet, momenta
 $\bvec{q}$ here correspond to a one-site unit cell. They can be
 obtained from momenta $\bvec{k}$ of the two-site unit cell as
 $q_{x/y} = (k_{a} \pm k _{b})/2$. The lines in the floor correspond
 to the mean-field (MF) bands for a Hubbard model with $t=1$, $t'=-0.5\;t$,
 and $U=10t$. Panel
 (c) shows the spectral density for spin up (blue) and down (red) at momentum
 $\bvec{q}=(\pi/2,\pi/2)$ and compares the MF approximation (plotted upwards) to the SCBA (plotted downwards).
 \label{fig:specs_checker} }
\end{figure*}

Figure~\ref{fig:specs_checker} shows the resulting one-particle spectrum for
$J'=0.15\;t$ and $t'=- 0.5\;t$ and compares it to mean-field (MF)
bands. 
In both approaches -- and especially around momenta $(\tfrac{\pi}{2},\pm \tfrac{\pi}{2})$ -- spectra clearly depend on spin both in MF and SCBA.
However, as shown in detail in (c), the spin dependence
manifests itself very differently: while MF theory predicts a
second peak of opposite spin
and at significantly different energy, 
this peak is largely lost within the
incoherent part of the spectrum in the SCBA. 
Instead, we  find at strong coupling 
a single coherent QP with 
spin-dependent spectral weight rather than spin-split bands.

\begin{figure}
  \includegraphics[width=0.49\columnwidth,clip]{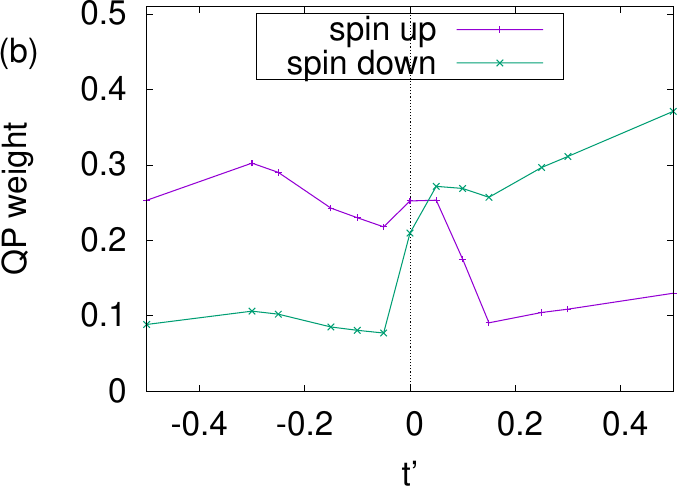}
  \includegraphics[width=0.49\columnwidth,clip]{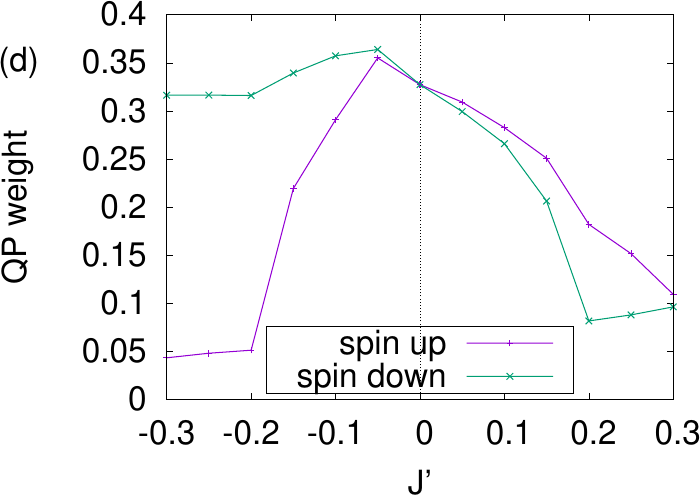}\\
  \caption{QP weights~\cite{SM} at momentum  
    $(\tfrac{\pi}{2},\tfrac{\pi}{2})$ for (a) $J'=0.15\;t$ and variable $t'$ and
    (b) $t'=0$ and variable $J'$.
   \label{fig:QP_weights}}
\end{figure}

Very close to the AF limit, i.e. for small $|t'|\ll t$, a second QP with the opposite dominant spin is seen at very slightly different energy, see the SM~\cite{SM}. However, as soon as larger $|t'|$ increases this energy splitting, the second QP reaches the continuum states, broadens and thus looses its QP character. The associated QP weight then drops sharply, see Fig.~\ref{fig:QP_weights}(a), and total QP weight becomes spin polarized.
%
$J'$ has a weaker effect on the bands than $t'$, so that two QPs of opposite spin can be found up to quite substantial values of
$|J'|\lesssim 0.2\;t = 2J/3$, see Fig.~\ref{fig:QP_weights}(b). However, their energy difference remains small throughout~\cite{SM}.

\begin{figure}
\includegraphics[height=0.36\columnwidth,trim= 0 0 80 0, clip]{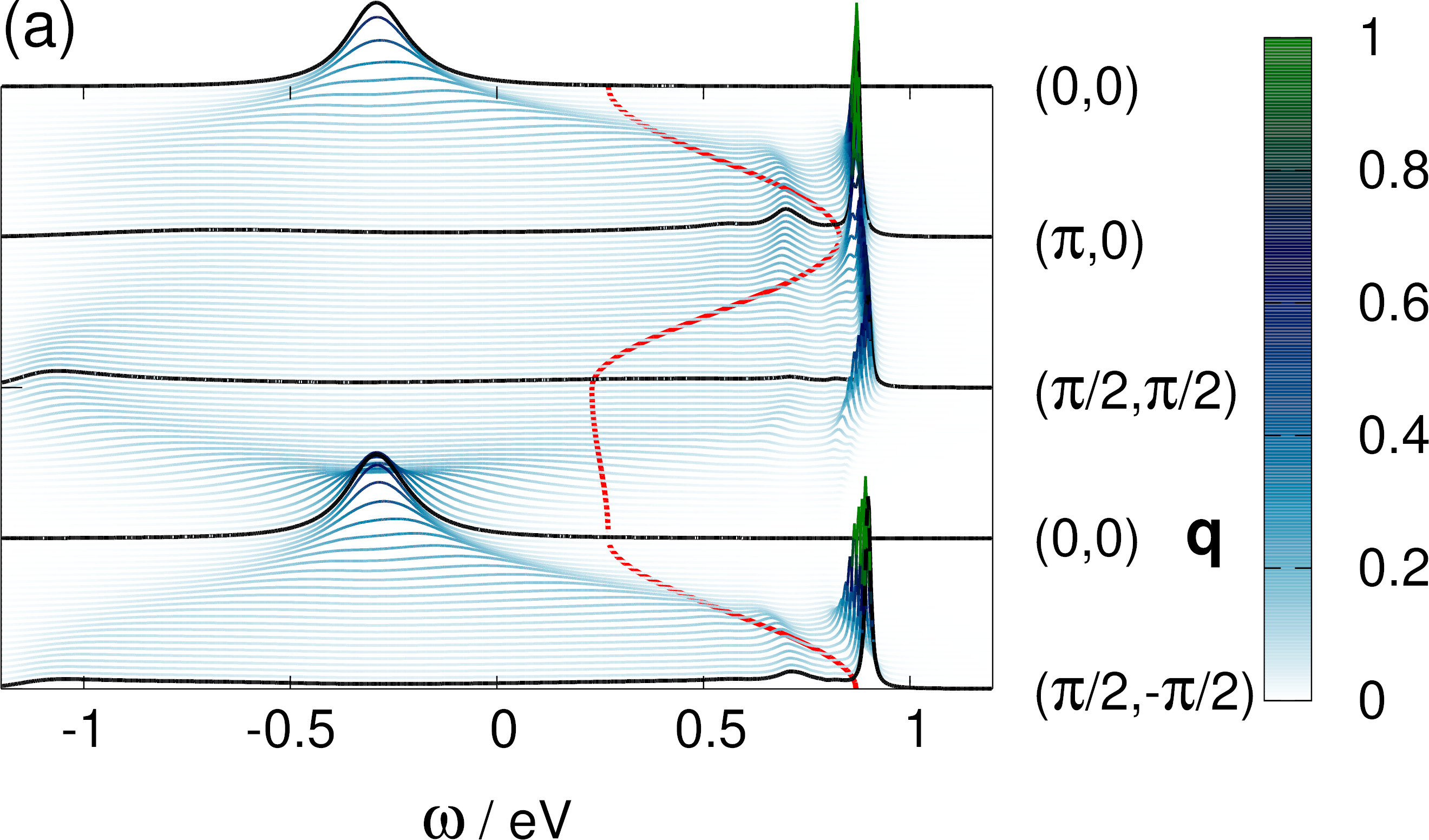}
\includegraphics[height=0.36\columnwidth,trim= 0 0 35 0,
  clip]{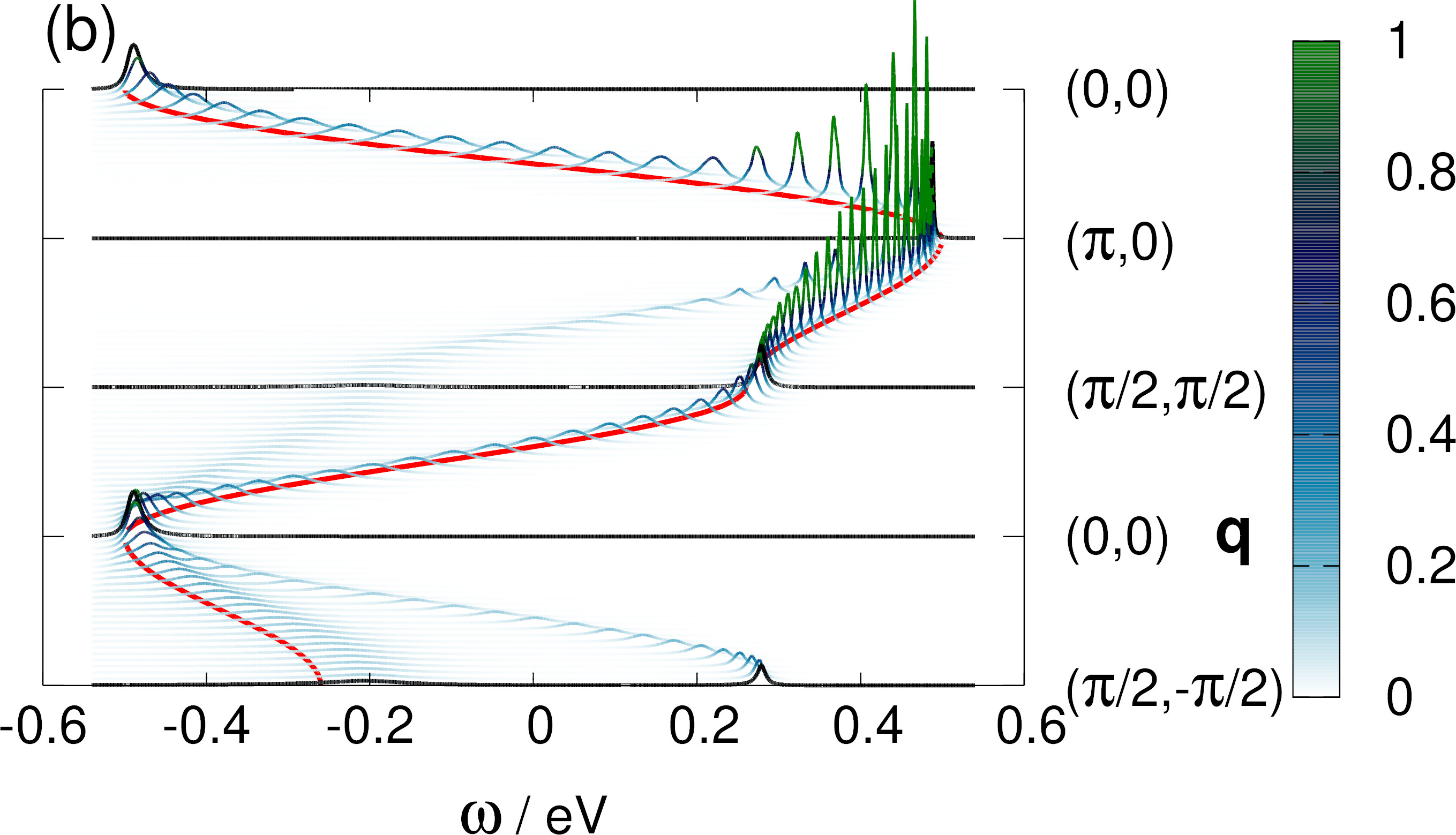}
\caption{One-particle spectral density obtained with SCBA for spin down
  on the generalized checkerboard 
  (representing transition-metal ions on the ILL)
  with $J = 10\;\textrm{meV}$, $J' =
  2.5\;\textrm{meV} \approx J'_b = 3\;\textrm{meV}$, and $|S|=\tfrac{5}{2}$.  In
  (a), NN hopping $t=300\;\textrm{meV}$, $t' = 150\;\textrm{meV} =
  \tfrac{t}{2}$, and $t'_b=10\;\textrm{meV}$ describe 
  an $xy$ orbital. (b) is for the $3z^2-r^2$
  orbital with
  $t=t'=60\;\textrm{meV}$ and  $t'_b=180\;\textrm{meV} = 3t$~\cite{2025arXiv250621661G}. Lines on the floor correspond to 
  the large-$U$ limit of MF, where the hole only hops on the spin-down sublattice.
  \label{fig:scba_xy_z2}}
\end{figure}

The transition-metal ions of the ILL, which hosts a number of
  altermagnets~\cite{2025arXiv250804839C,s31h-hk2v}, form a  generalization of the 
  checkerboard lattice, where additional couplings 
  $t'_b$ and $J'_b$ are introduced along the dotted diagonals of
  Fig.~\ref{fig:cartoon_checker}. In the Mott 
  insulator La$_2$O$_3$Mn$_2$Se$_2$, for instance, the two 
  exchange couplings $J'_b\approx J'$ between NNN Mn ions happen to be rather similar, but band structure
  clearly indicates quite  different NNN hoppings $t'_b\neq
  t'$~\cite{Wei24,2025arXiv250621661G}.
   Hoppings $t'_b \ll t'\approx
  \tfrac{t}{2}$ for the Mn $xy$ orbital give
  Fig.~\ref{fig:scba_xy_z2}(a) showing a single QP with
  spectral-weight transfer, as above in Fig.~\ref{fig:specs_checker}. As a second
  case, we look at the Mn $3z^2-r^2$ orbital, which has small NN hopping $t$, so
  that $t'\approx t$ and $t'_b \approx 3t$. Sublattice-conserving hoppings thus dominate, the hole
  decouples from the spin background and the two spin-split bands survive to a significant
  extent, see Fig.~\ref{fig:scba_xy_z2}(b). Note, however, that $t_{2g}$ and $e_g$
   bands are expected to hybridize and overlap in
   La$_2$O$_3$Mn$_2$Se$_2$, so that measured spectra would show a mixed character.

We have complemented the SCBA study with the variational cluster 
approximation (VCA)~\cite{Pot03b,Pot03b} applied to the checkerboard Hubbard model. The $J$ and $J'$ terms of
(\ref{eq:ham_checker_J}) are here replaced
by an onsite Hubbard repulsion $U\sum_i n_{i,\uparrow}n_{i,\downarrow}$. We choose $U=10\;t$ which gives an effective NN superexchange $J=\tfrac{4t^{2}}{U}=0.4\;t$ as before, and puts us solidly
in the Mott insulating regime at half filling. Further setting $t'=\pm 0.2\;t$ and $\pm 0.5\;t$, we then verified~\cite{SM} that
the magnetically ordered state has a QP
with momentum-dependent spin polarization, analogous to the SCBA
spectra discussed in Fig.~\ref{fig:specs_checker} above.


\begin{figure}
  \includegraphics[height=0.48\columnwidth,trim=0 0 0 0,clip]{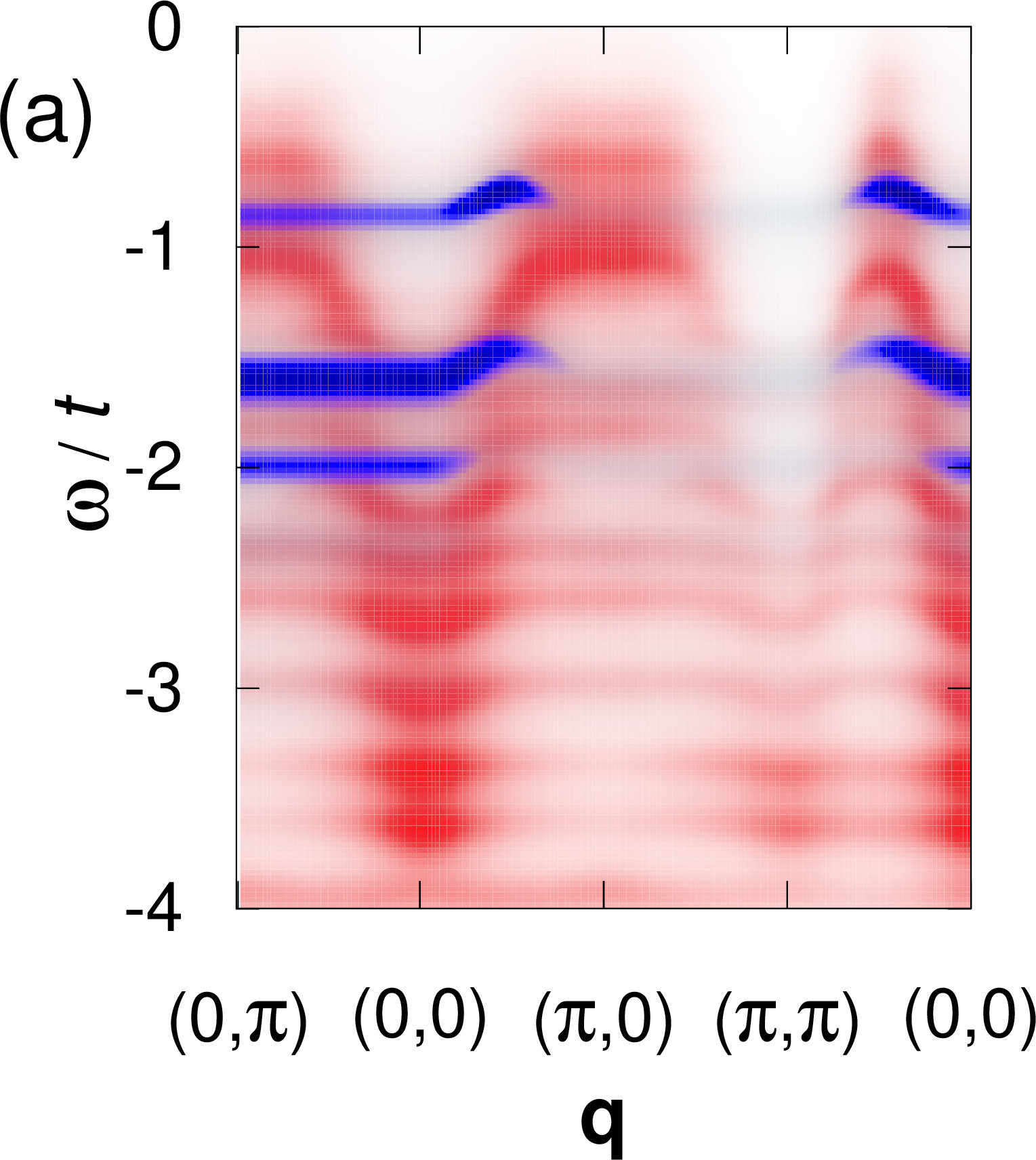}
  \includegraphics[height=0.48\columnwidth,trim=0 0 0 0,clip]{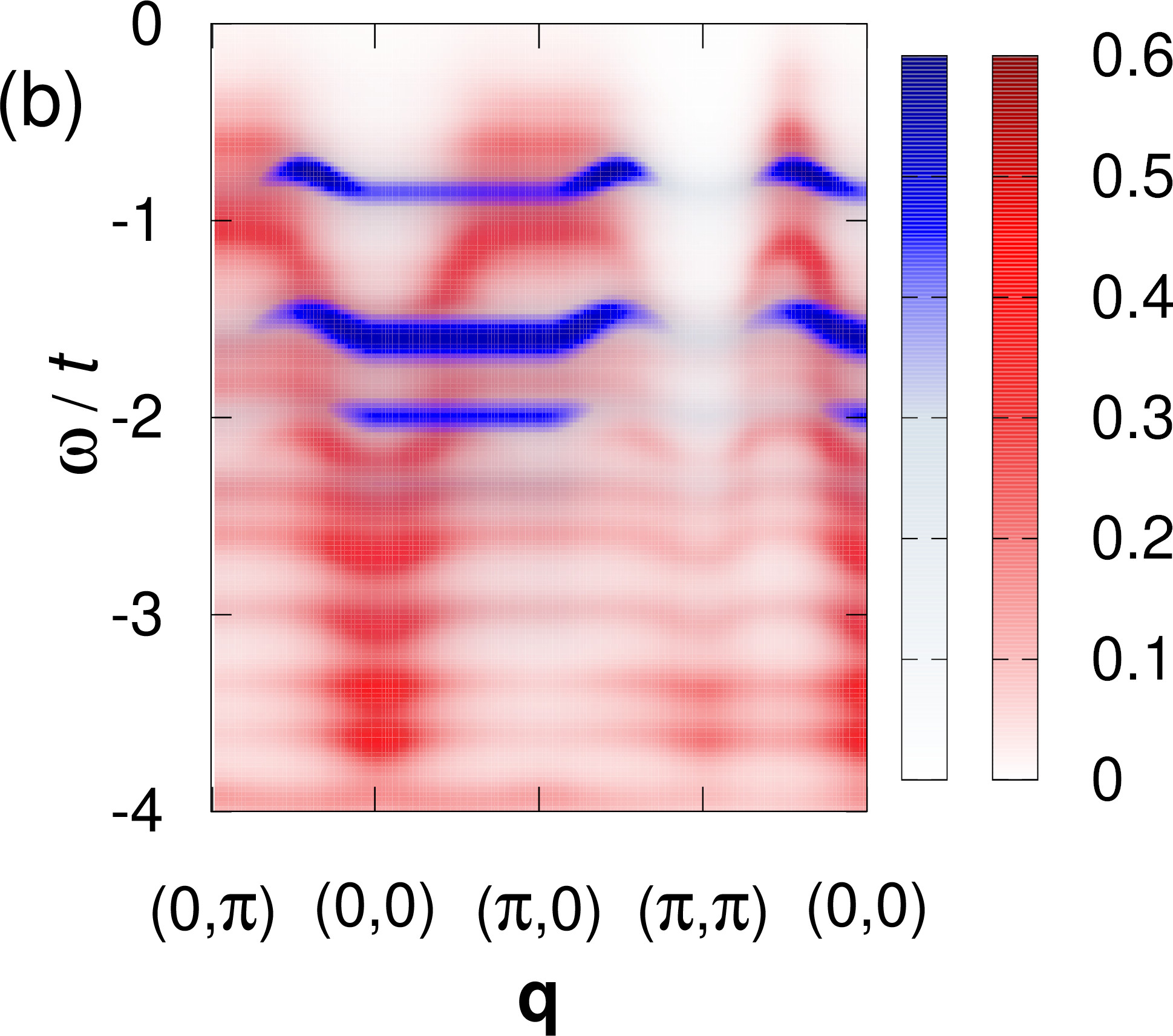}\\
  \includegraphics[height=0.42\columnwidth,trim=0 0 0 0, clip]{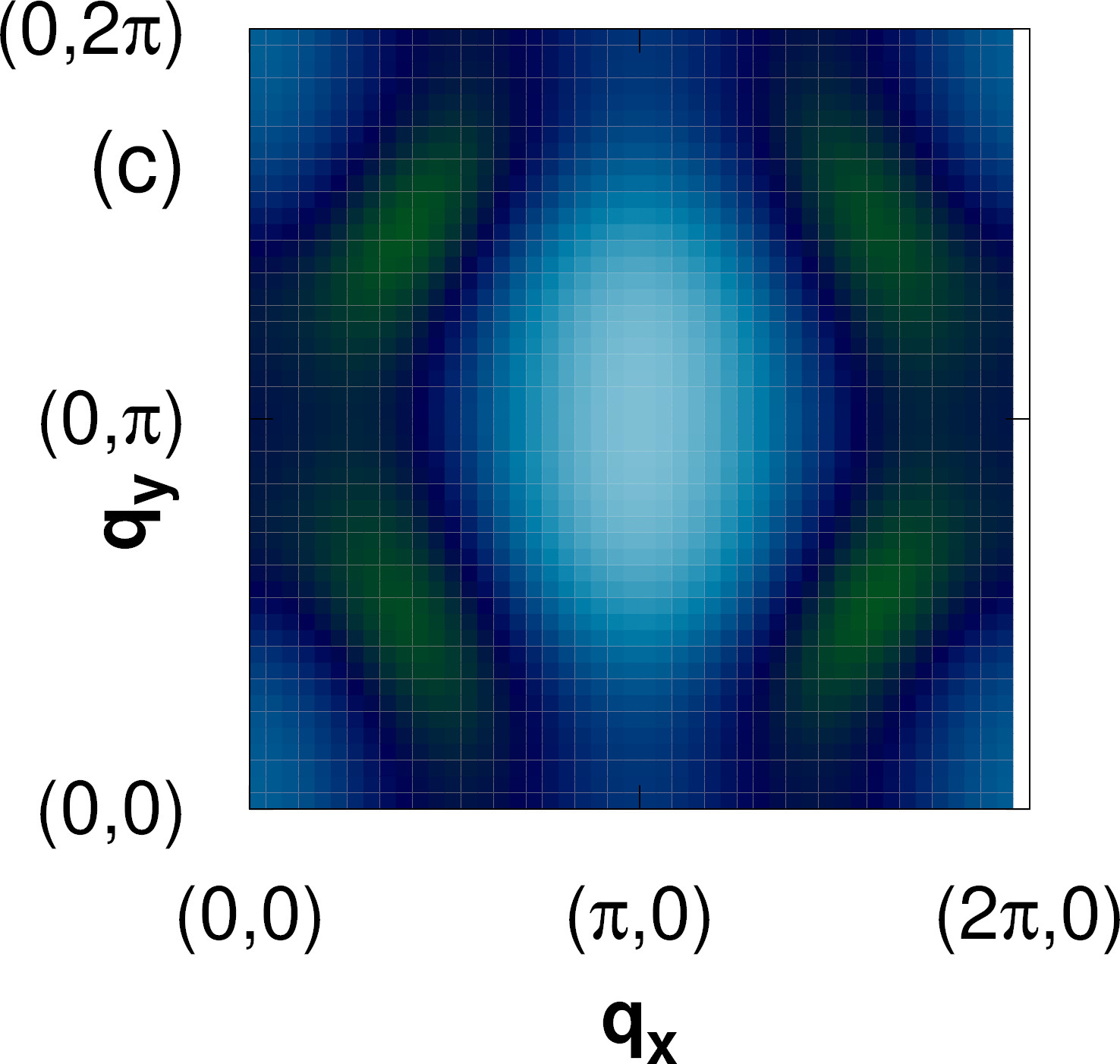}
  \includegraphics[height=0.42\columnwidth,clip]{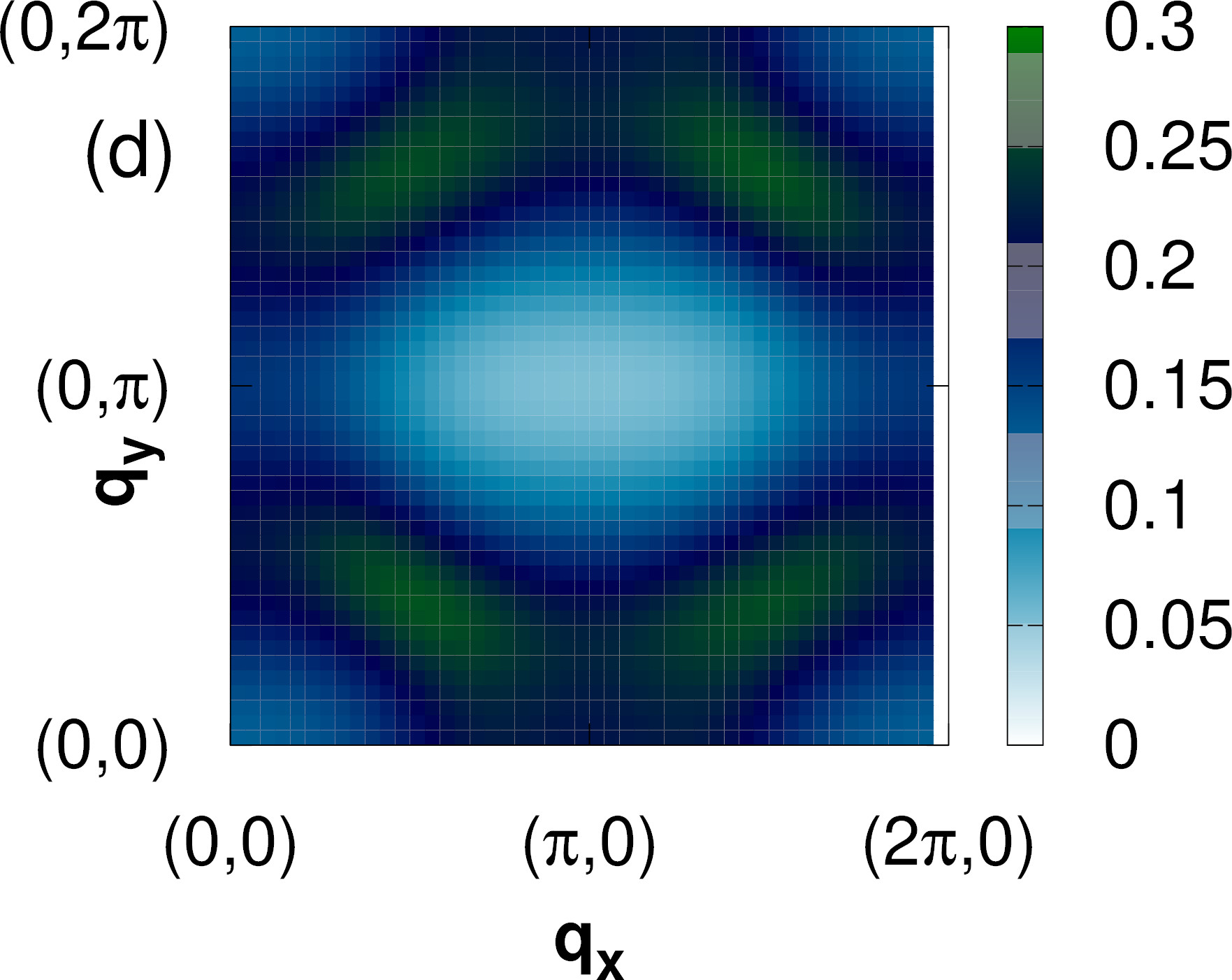}
  \caption{Occupied states obtained with VCA for the
    three-orbital $t_{2g}$ model with $t=1$, $U=14 t$, $J_H=2 t$,
    $\Delta=-0.5\;t$ in
    the AF and AO state illustrated in Fig.~\ref{fig:cartoon_LaVO3}. (a) gives the energy- and momentum dependent
    spectrum for spin up and (b) for spin down. Weight in red comes from the $xy$ orbital, while the combined $xz$ and $yz$ weight is shown in blue.  (c)
    and (d) show the weight of up. resp. down states integrated over the energy
    range $-1.2 t< \omega < -0.5 t$, all orbitals.\label{fig:specs_t2g}}
\end{figure}

\textit{Altermagnetic polaron spectral function in LaVO$_{3}$ ---} 
To establish a platform where the altermagnetic polaron can be experimentally
identified and investigated, we propose layers or surfaces of LaVO$_{3}$. Bulk LaVO$_{3}$ has $G$-type orbital and $C$-type spin order, which yields an AM state for single planes.
We apply the VCA to a two-dimensional three-orbital $t_{2g}$ model with kinetic energy
\begin{align}
  H_{t_{2g}} = -t\sum_{\langle i,j\rangle,\sigma }
  \cdag_{i,xy,\sigma}\cnod_{j,xy,\sigma}
  +\Delta \sum_{i,\sigma }n_{j,xy,\sigma}\\
  - t \sum_{\langle i,j\rangle\parallel a,\sigma } \cdag_{i,xz,\sigma}\cnod_{j,xz,\sigma}
  - t \sum_{\langle i,j\rangle\parallel b,\sigma } \cdag_{i,yz,\sigma}\cnod_{j,yz,\sigma}
\end{align}
on the square lattice, 
where $\cdag_{i,\alpha,\sigma}$ refers to an electron with spin $\sigma$ in orbital
$\alpha$ on site $i$. 
The dominant energy scale is here onsite interaction $U$ together with Hund's-rule
coupling~\cite{SM}, so that a filling of two electrons corresponds to a
Mott insulator with large spin one. Crystal field splitting $\Delta < 0$
ensures that one electron is found in the $xy$ orbital, and this
half-filled $xy$ orbital strongly promotes AF magnetic order. The
second electron can be in either the $xz$ or the $yz$ state and the
VCA here finds the AF/AO order of Fig.~\ref{fig:cartoon_LaVO3} to be
favorable. 


The orbital order has a strong Ising-like -- rather than
Heisenberg-like -- character, and the AF/AO state transfers this 
to the spin degree of freedom as well~\cite{PhysRevB.79.224433}.
The VCA spectra for the AF/AO state are shown in Fig.~\ref{fig:specs_t2g} and
show the QP peaks with spin-dependent dispersion, as expected from earlier
work~\cite{PhysRevB.79.224433}.
Hole motion is almost completely driven by three-site
hopping of the hole~\cite{PhysRevLett.100.066403,PhysRevB.78.214423} that acts along the
$x$ ($y$) axis for a hole inserted into the $xz$ ($yz$) orbital. As spin and orbitals are locked
together by the AO/AF order, a hole inserted into a particular orbital also
has a definite spin polarization. The spin-dependent dispersion is then repeated in
a few replica bands with increasing excitation energy (`ladder spectrum'),
which are due to the polaronic motion in an Ising-like AO/AF
state.

In addition to spin-dependent dispersion, however, we also find substantial
spin-dependent SW transfer---see spectra at the special AM
momenta as well as the integrated SW in Fig.~\ref{fig:specs_t2g}(c) and
(d): Spectra around momenta $\bvec{q} = (\pi,0)$ and $(0,\pi)$ are strongly
spin polarized, especially within the combined $xz$/$yz$ sector. (In the $xy$
orbital, both spins can hop in both directions, which somewhat dilutes
polarization.) 
SW is     
here not transferred into higher-energy incoherent features, all peaks
of the ladder spectrum in contrast have a similar weight
distribution. Instead, the interplay between polaronic and free holon motion
moves some weight of opposite spin into the upper Hubbard band.  

\textit{Conclusions and outlook ---} We have here extended the discussion of altermagnetic
symmetries to the strong-coupling case of a spin polaron formed by holes in
altermagnetic Mott insulators. Spin-split QP peaks of both spin
species, as would be expected at weak coupling, may be
preserved when coupling to spin fluctuations is suppressed. This can 
happen in the case of an Ising spin background, as discussed here for a
$t_{2g}$-model that could be realized in surfaces or layers of orbitally
ordered LaVO$_{3}$.  Nevertheless,  weights of the two QPs are even here found to be quite
different, resulting in a modification of the weak-coupling picture and in effective spin-momentum locking of the QP
sector. Similarly, when a hole in the $3z^{2}-r^{2}$ 
orbital of the ILL Mott altermagnet
La$_2$O$_3$Mn$_2$Se$_2$~\cite{Wei24,2025arXiv250621661G}
moves mostly on one -- largely spin-polarized -- sublattice, two spin-split QP
are expected. If the hole instead 
couples more strongly to spin fluctuations, as in the checkerboard model or the
$xy$ orbital of La$_2$O$_3$Mn$_2$Se$_2$, the two spin-polarized QPs are replaced by a single QP peak with clear
spin-momentum locking, while the spectral weight of
opposite spin is incoherent and broad.

The effect of altermagnetic band splitting on superconductivity has
already become an extensive area of research. For instance, the band splitting
can give Cooper pairs a finite momentum even without a magnetic
field~\cite{PhysRevB.110.L060508,pub.1169203882}, one may think of a
superconducting diode effect~\cite{PhysRevB.110.024503}, and the Lieb
lattice formed by Cu and O ions in cuprate superconductors can form an AM~\cite{2024arXiv241211922L}.
In this context our findings raise the question of
what happens when stronger correlations not only shift the band
energies, but instead render bands of one spin species
incoherent. Indeed, QP coherence has recently been found to have a
  signifcant impact on superconducting instabilities~\cite{2025arXiv251006313S,2025arXiv250925318S}.

This point immediately relates to a longstanding debate in the field of
high-temperature superconductors on pairing mediated by correlations~\cite{Anderson_glue_07}. 
The spin polaron suppresses
spin order in the vicinity of the mobile charge, which self-consistently
traps the charge~\cite{Schrieffer1988}. This so-called spin bag does not carry a spin in
antiferromagnets, but from our results 
it is clear that in altermagnets it does and it will be interesting to establish how this affects the
pairing mediated by a shared bag. Very
recently, numerical studies of the square-lattice Hubbard model have addressed
the relative importance of pairing mechanisms due to spin fluctuations
vs. others based more directly on short-range 
interactions~\cite{Dong_DCA_22a,Dong_DCA_2022,Yamase_2023}. The
difference between AM and AF when it comes to
polaron spin and spin-momentum locking suggest that numerical
 investigations of superconductivity in strongly correlated
altermagnets, beyond MF~\cite{PhysRevB.111.104432,s31h-hk2v}, may here contribute
further insights, as might experimetnal realizations based on  strongly-correlated building
blocks~\cite{PhysRevLett.134.166701}.

{\it Acknowledgments ---}
We thank Johannes Knolle for helpful disussion. 
K.W. thanks National Science Center, Poland for financial support 
(grant number 2024/55/B/ST3/03144).
Part of this work was supported by the Deutsche Forschungsgemeinschaft (DFG,
German Research Foundation) through the Sonderforschungsbereich SFB 1143,
grant No. YE 232/2-1, and under Germany's Excellence Strategy through the
W\"{u}rzburg-Dresden Cluster of Excellence on Complexity and Topology in
Quantum Matter -- \emph{ct.qmat} (EXC 2147, project-ids 390858490 and 392019).  

{\it Note added---}
While working on this manuscript, we became aware of work by
L. {L}anzini \textit{et al.} discussing related issues~\cite{Lanzini}.


%

\newpage

\renewcommand{\thefigure}{S\arabic{figure}}
\renewcommand{\theequation}{S\arabic{equation}}
\renewcommand*{\citenumfont}[1]{S#1}
\renewcommand*{\bibnumfmt}[1]{[S#1]}

\setcounter{equation}{0}
\setcounter{figure}{0}

\section*{Supplemental Material: Altermagnetic polarons}
This supplemental material contains details on the self-consistent
Born approximation (SCBA) and the variational cluster approximation,
as well as on the $t_{2g}$ Hamiltonian based on LaVO$_3$.

\subsection{Self-consistent Born approximation}
This section contains for completeness the equations used in obtaining the
results with SCBA presented in the main text.

Equation (3) of the main text is obtained via 
linearised spin wave,
Fourier and Bogoliubov
transformations~\cite{Martinez1991}. After Fourier transform, a spin flip away from the N\'eel reference state is
denoted by operators
\begin{align}
  a_{\bvec{k}} = u_{\bvec{k}}\alpha_{\bvec{k}} - v_{\bvec{k}}\beta_{-\bvec{k}}^{\dagger}\\
  b_{\bvec{k}} = u_{\bvec{k}}\beta_{\bvec{k}} - v_{\bvec{k}}\alpha_{-\bvec{k}}^{\dagger}\;,
\end{align}
where $a_{\bvec{k}}$ and $b_{\bvec{k}}$ refer to a spin flip on $A$ and $B$
sublattices, resp. Coefficients $u_{\bvec{k}}$ and $v_{\bvec{k}}$ are chosen to diagonalize the
magnon Hamiltonian
\begin{align}
  &\left(\adag_{\bvec{k}},\bnod_{\bvec{k}}\right)
    \left(\begin{array}{cc}
            h_A      & h_{AB}\\
            h_{AB} & h_B
          \end{array}\right)
                     \left(\begin{array}{c}
                             \anod_{-\bvec{k}}\\\bdag_{-\bvec{k}}
                           \end{array}\right)=
  \omega_{\alpha,\bvec{k}} \alpha_{\bvec{k}}^\dagger\alpha_{\bvec{k}}^{\phantom{\dagger}}
  +\omega_{\beta,\bvec{k}}
  \beta_{\bvec{k}}^\dagger\beta_{\bvec{k}}^{\phantom{\dagger}}
\end{align}
with 
\begin{align}
  h_{A/B} &= 2S(2J-J'  + J'\cos k_{a/b}),\ 
          h_{AB}=4S J \gamma_{\bvec{k}}\;,\\
  \gamma_{\bvec{k}}&\!=\!\tfrac{1}{2}(\cos \tfrac{k_{a}+k_{b}}{2} +
\cos \tfrac{k_{a}-k_{b}}{2})
  \!=\!\tfrac{1}{2}(\cos q_{x} +\cos q_{y})\;,
\end{align}
where $S=\tfrac{1}{2}$ refers to the electron spin. 
Momenta $\bvec{k} = (k_{a},k_{b})$ refer to the first Brillouin zone of the
checkerboard lattice with its two-site unit cell, momenta $\bvec{q} =
(q_{x},q_{y})$ would correspond to a square lattice with a single-site unit
cell. One then finds 
\begin{align}\label{eq:bogo}
  u_{\bvec{k}}& = \cosh \theta,\  v_{\bvec{k}}= \sinh\theta,\ \theta = \frac{1}{2}\atanh\left(\frac{2 h_{AB}}{h_A+h_B}\right)
\end{align}
and
\begin{align}
  \omega_{\alpha/\beta,\bvec{k}} &= h_{A/B} v_{\bvec{k}}^2 + h_{B/A} u_{\bvec{k}}^2 - 2u_{\bvec{k}}v_{\bvec{k}}h_{AB}\;.
\end{align}
This completes the transformation of the spin Hamiltonian (2) into the Eq. (3)
of the main text.

Next, using successive slave-fermion, Fourier and Bogoliubov  transformations, see~\cite{Martinez1991} for details, we express kinetic energy, Eq.~(1) of the main text
in terms of hole $h$ and,
if needed, magnon $\alpha, \beta$ creation
and annhiilation operators.
NNN electron hopping within a sublattice
leads to hole moving without 
coupling to magnons and reads
\begin{align}\label{eq:H_t2_eff}
  \tilde{H}_{t,2} = 2t'\sum_{\bvec{k}}\left( \cos
  k_{a}\; \hdag_{\bvec{k},A}\hnod_{\bvec{k},A}
  +\cos k_{b}\;\hdag_{\bvec{k},B}\hnod_{\bvec{k},B}\right)\;,
\end{align}
where $\hdag_{\bvec{k},\alpha}$ ($\hnod_{\bvec{k},\alpha}$) creates
(annihilates) a hole with momentum $\bvec{k}$ on sublattice $\alpha=A,B$. 
NN hopping between sublattices in contrast creates or annihilates magnons and thus couples hole motion and
magnons:
\begin{equation}\label{eq:H_t1_eff}
  \tilde{H}_{t}\!=\!\frac{zt}{\sqrt{N_{u}}}\!\sum_{\bvec{k},\bvec{k}'} \! M_{\bvec{k},\bvec{k}'} \!\left(
      \hdag_{\bvec{k},A}\hnod_{\bvec{k}-\bvec{k}',B}\alpha_{\bvec{k}'}  + 
  \hdag_{\bvec{k},B}\hnod_{\bvec{k}-\bvec{k}',A}\beta_{\bvec{k}'}\right) + \textrm{H.c.},
\end{equation}
with $z=4$ NN bonds per site and the number $N_{u}$ of two-site unit cells. 
$M_{\bvec{k},\bvec{k}'} = \left(-v_{\bvec{k}'}\gamma_{\bvec{k}}+u_{\bvec{k}'}\gamma_{\bvec{k}-\bvec{k}'}\right)$
depends on the constants $u_{\bvec{k}'}$ and $v_{\bvec{k}'}$ arising in the
Bogoliubov transformation (\ref{eq:bogo}) above. Equations~(\ref{eq:H_t2_eff})
and ~(\ref{eq:H_t1_eff}) are the one-hole approximation of the kinetic energy
(1) of the main text.

We are here interested in spectral densities obtained from sublattice-dependent hole Green's functions
\begin{align}
  G^{\alpha}(\bvec{k},\omega) = \langle
  \phi_{0}|\hnod_{\bvec{k},\alpha}\frac{1}{\omega -(\tilde{H} - E_{0}) +i\delta} \hdag_{\vec{k},\alpha}|\phi_{0}\rangle\;,
\end{align}
where $\tilde{H} = \tilde{H}_{J}+\tilde{H}_{t}+\tilde{H}_{t,2}$ is
the approximated one-hole Hamiltonian, $\alpha=A, B$ is the sublattice index, and the ground state $|\phi_{0}\rangle$ is the undoped
linear--spin-wave ground state. Dyson equation and self energy of the main
text then complete the SCBA for the sublattice-dependent Green's functions. 
Alternating magnetic order ties spin character to
sublattice, however, quantum fluctuations reduce the correlation so that 
\begin{align}
  G^{\uparrow}(\bvec{k},\omega) = G^{A}(\bvec{k},\omega)
  +\left(\frac{1}{N_{u}}\sum_{\bvec{k}} v_{\bvec{k}}^{2}\right)
  \left(G^{B}(\bvec{k},\omega)-G^{A}(\bvec{k},\omega)\right)
\end{align}
and analogously for $G^{\downarrow}(\bvec{k},\omega)$. These corrections can
become quite substantial for large $J'$.

For the Lieb-lattice calculations giving Fig. 4 of the main text, we
proceed analogously, which modifies
\begin{align}
  h_{A/B} &= 2S(2J- J'  + J'\cos k_{a/b}- J'_b  + J'_b\cos k_{b/a})
  \quad\textrm{with}\quad
  S=\frac{5}{2}\;.
\end{align}
The free dispersion for sublattice $a/b$ becomes
\begin{align}
  \epsilon^{a/b}(\bvec{k}) = 2t'\cos k_{a/b} + 2t'_b\cos k_{b/a}\;.
\end{align}

\subsubsection{Extraction of QP weight}

\begin{figure}
  \includegraphics[width=0.8\columnwidth,clip]{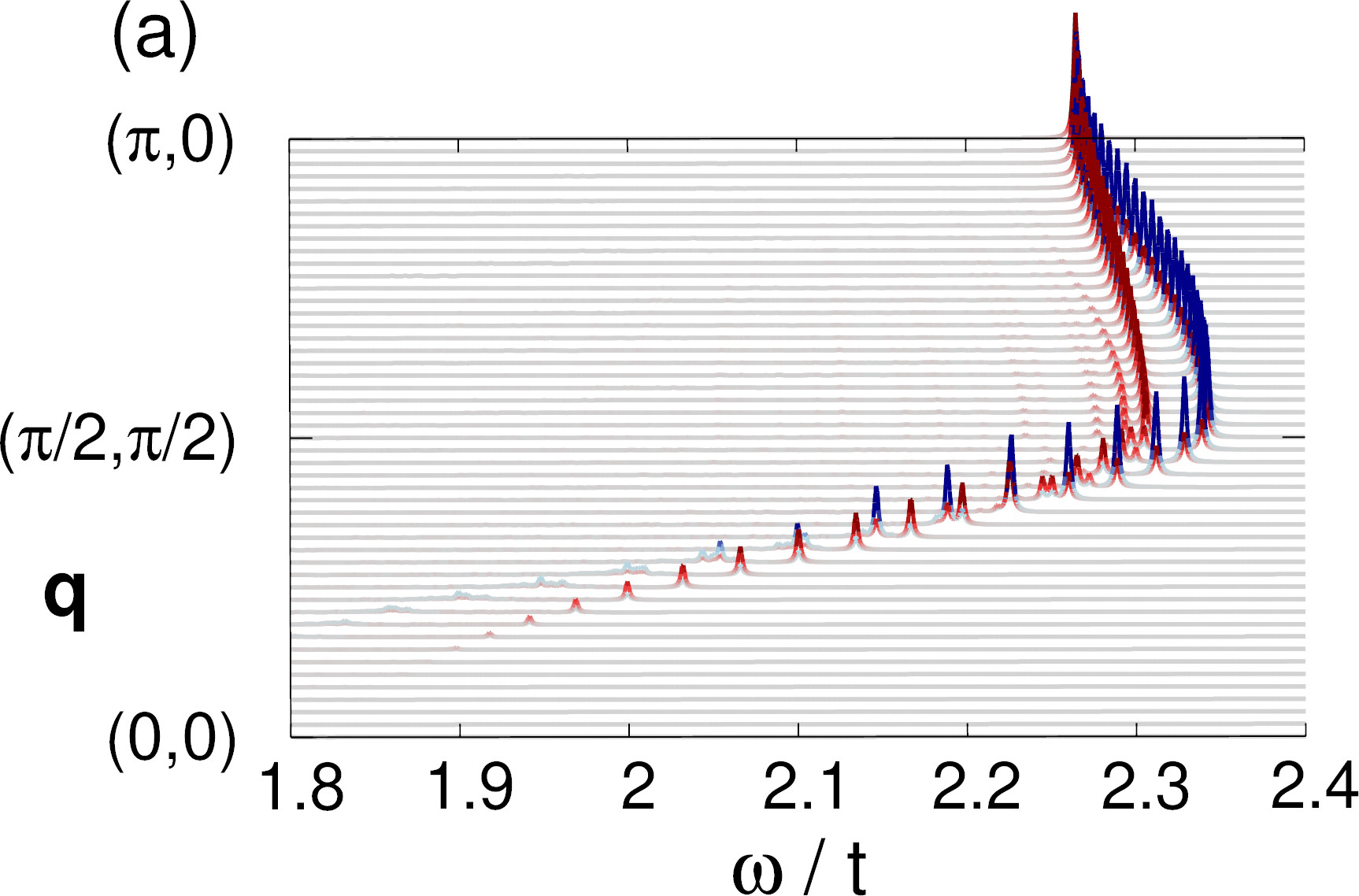}\\
  \includegraphics[width=0.8\columnwidth,clip]{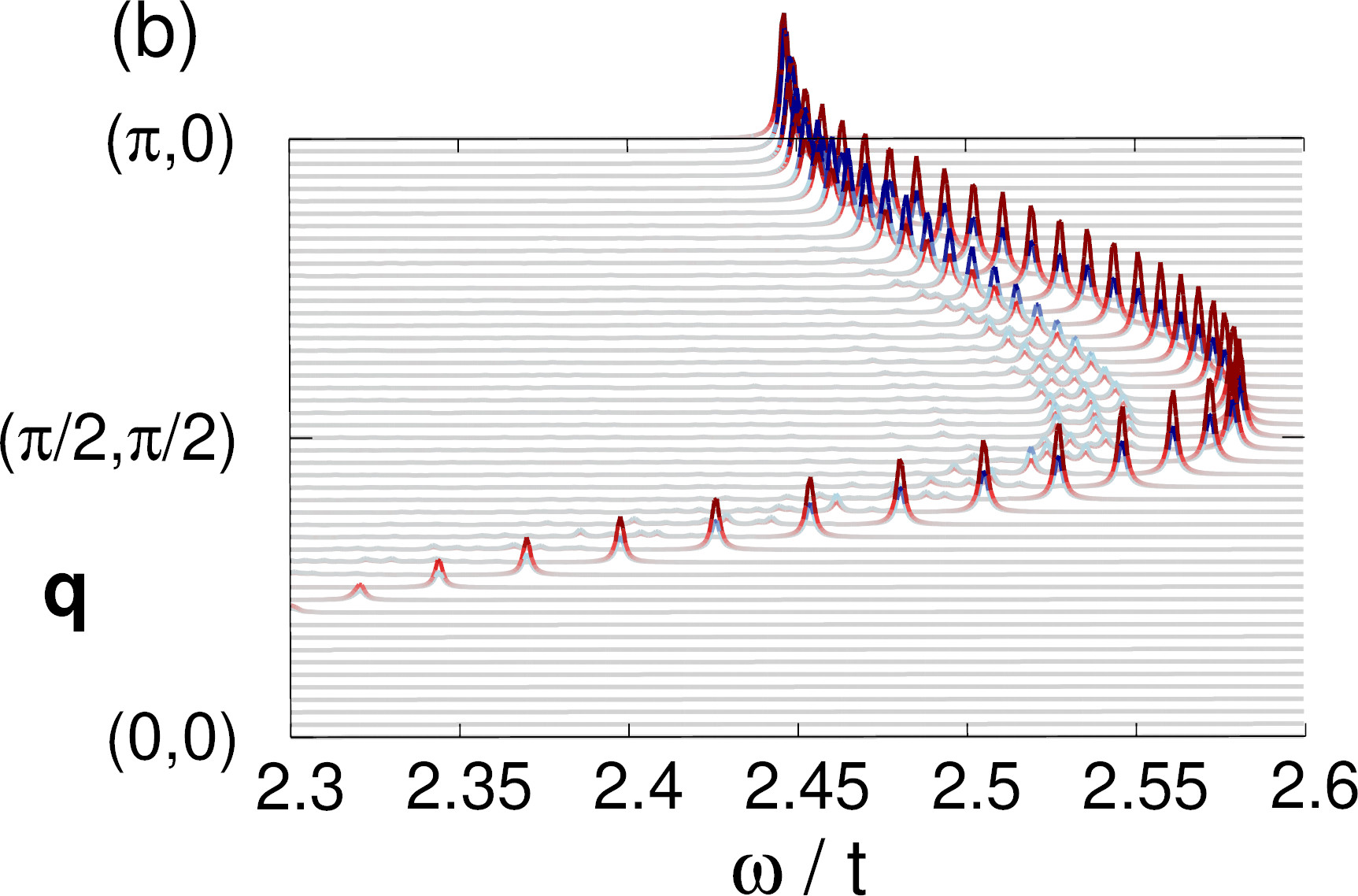}
  \caption{High-resolution spectra for spin up (red)
    and down (blue) and parameters (a) $J'=0.15\;t$, $t'=0.1\;t$ and
    (b) $J'=0.25\;t$, $t'=0$. 
   \label{fig:high_res_specs}}
\end{figure}

In the $t$-$J$ model at half filling, the integral over the occupied
sepctrum adds to one at each momentum and for each spin. Without band
splitting, the 
QP weight can be taken to be the fraction of the weight coming from the
top-most peak. Here, however, we have to take into account the
possibility of there being two QPs at the momenta
$\vec{q}=(\tfrac{\pi}{2},\pm \tfrac{\pi}{2})$ that we are interested
in. We thus use the following scheme:
\begin{itemize}
  \item We use very high energy resolution that could not be
    realistically expected from ARPES experiments to resolve the tiny
    altermagnetic band splitting.
  \item
    The QP part of the spectrum always includes the
    lowest-energy excitation, which is the highest peak and well
    separated from the remaining spectrum for all parameter sets.
  \item 
    For some parameter sets (namely for $t'\ll t$), a second peak (of opposite spin polarization to
    the first) is also separated from the remaining spectrum, it was
    then included into the QP weight. An example is shown in
    Fig.~\ref{fig:high_res_specs}(a), where the two right-most peaks
    are significantly larger that all others, and can be clearly identified
    over a large momentum range. 
  \item
    For larger $t'$ or large $J'$, only one peak can be clearly separated
    from all other spectral features, the spectrum is here given by
    this peak and a large number of small peaks, which would in the
    thermodynamic limit be expected to merge into a continuum. This is
    seen in Fig.~\ref{fig:high_res_specs}(b): After the first peak, we
    see a number of clustered features of comparable weight.
  \item
    The criterion to decide whether to consider the second excitation
    peak as a second QP or as the first peak of the continuum was
    wether this second peak is larger or smaller than the third peak.
  \item We verified that changing the threshold for the comparison
    does not significantly affect the outcome: Once the second peak
    reaches the continuum, it soon broadens and merges into it. 
\end{itemize}

\subsection{Variational Cluster approximation}
\begin{figure}
  \includegraphics[width=0.49\columnwidth]{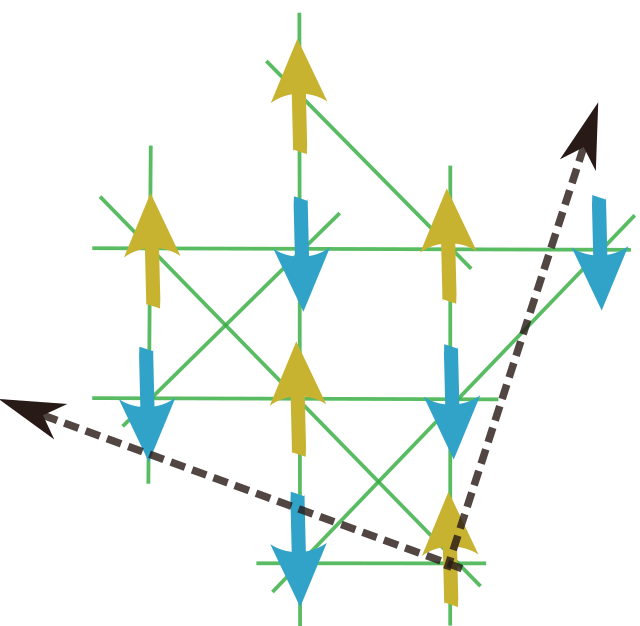}
  \caption{Ten-site cluster that was used in the variational
    cluster approximation for the checkerboard-lattice Hubbard model. Black arrows show how the clusters are periodically connected.\label{fig:clusters} }
\end{figure}

In addition to the SCBA applied to the
$t$-$J$ model we also study a Hubbard model on the checkerboard lattice
with the  variational cluster approximation
(VCA)~\cite{Pot03b,Aic04}. As in cluster perturbation theory, the self
energy $\Sigma$ of the system is replaced by the self energy
$\Sigma_{\text{Cl}}$  of a small
cluster, which is in turn extracted from the Green's function
$G_\text{Cl}$ of the cluster.  The
cluster self energy $\Sigma_{\text{Cl}}$ is then combined with the
non-interacting Green's function $G^{-1}_0$ of a lattice to yield an
approximation
\begin{align}
  G = (G^{-1}_0 - \Sigma)^{-1} \approx (G^{-1}_0 - \Sigma_{\text{Cl}})^{-1}
\end{align}
of the full lattice Green's function.
We use here the ten-site cluster of Fig.~\ref{fig:clusters}, as it is more symmetric
than eight- or twelve-site clusters and the next highly symmetric
cluster (sixteen sites) would be numerically rather demanding.

In the VCA, the self energy is further optimized by finding a stationary
point of the grand potential w.r.t. one-particle parameters of the
Hamiltonian. The grand canonical potential $\Omega$ of the lattice system can
be written as 
\begin{equation}
  \Omega = \Omega_\text{Cl} + Tr\ln{[G^{-1}_0 - \Sigma_{\text{Cl}}]^{-1}} - Tr\ln{(-G_\text{Cl})},
  \label{eq:omega}
\end{equation}
where $\Omega_\text{Cl}$ is the grand canonical potential of the
cluster~\cite{Pot03a,Aic06b}. In our case, we find that the self energy calculated with an
additional small staggered magnetic field lowers the system's
energy. However, it must be emphasized that the calculated observables
(e.g. the spectral density) are obtained using the `original'
Hamiltonian, i.e., the staggered field is not included in the
non-interacting lattice Green's function $G^{-1}_0$.

\begin{figure}
  \includegraphics[width=0.49\columnwidth]{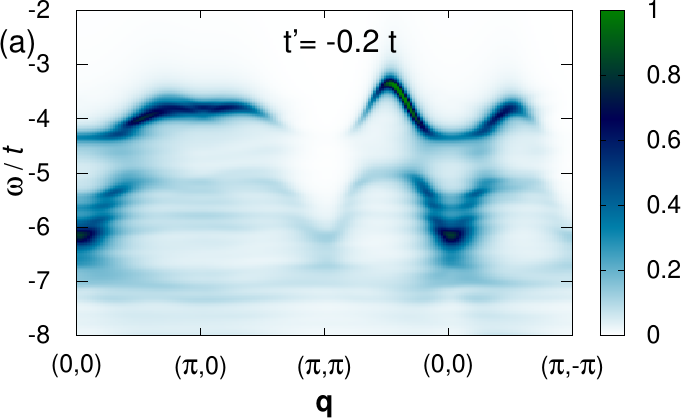}
  \includegraphics[width=0.49\columnwidth]{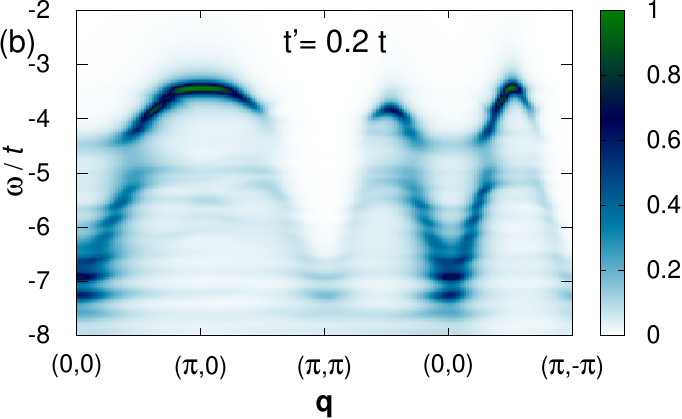}\\
  \includegraphics[width=0.49\columnwidth]{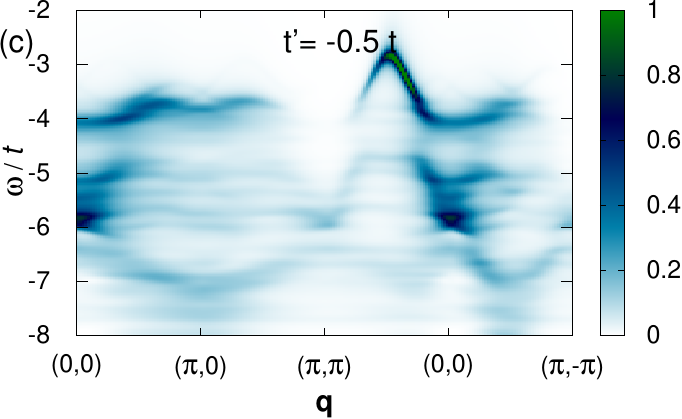}
  \includegraphics[width=0.49\columnwidth]{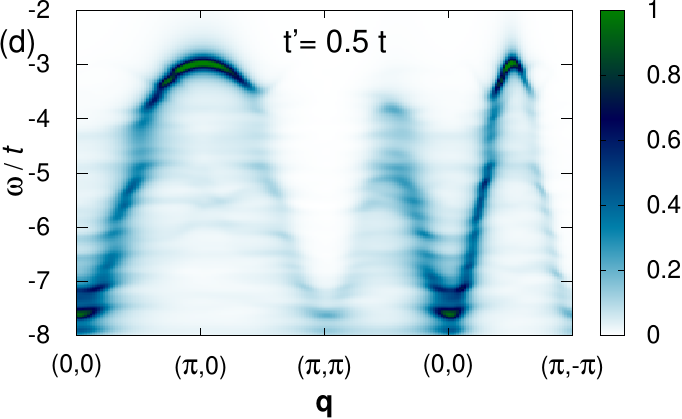}\\
  \caption{One-particle spectral density for spin up, obtained with VCA for the
    checkerboard model with $U=10\;t$ in the AF state
    optimizing the grand potential for (a) $t'=-0.2\;t$, (b)
    $t'= 0.2\;t$, (c) $t'=-0.5\;t$, and (d)
    $t'=0.5\;t$. 
    \label{fig:specs_checker_VCA}}
\end{figure}

Figure~\ref{fig:specs_checker_VCA} shows the one-particle spectra obtained in
the AF state optimizing the grand potential at half filling. The onsite
interaction $U$ was here set to $10\;t$, giving $J=\tfrac{4t^{2}}{U}=0.4\;t$
as in the main text, and second-neighbor hopping to
$t'=\pm 0.5\;t$ resp. $t'=\pm 0.2\;t$. The relevant momenta for
altermagnetism are
$\bvec{q}=(\tfrac{\pi}{2},\pm\tfrac{\pi}{2})$ resp. $\bvec{k}=(0,\pi)$
and $\bvec{k}=(\pi,0)$. While the spectrum for $t'=\pm 0.2\;t$ shows
one peak for each spin projection, one (closer to the Fermi level)
has much higher weight than the other and is also sharper. For the larger $t'=\pm 0.5\;t$,
only one coherent QP peak with a  clear spin polarization around these
momenta is seen, as in the SCBA results of the main text.

\subsection{Details of the $t_{2g}$ Hamiltonian}

Finally, we also looked at a three-orbital model for $t_{2g}$ orbitals, based
on LaVO$_{3}$. The onsite interactions are here taken 
\begin{align}
H_{\textrm{int}} &= U \sum_{i, \alpha} n_{i \alpha \uparrow} n_{i \alpha \downarrow} 
  +\frac{U^\prime}{2} \sum_{i, \sigma} \sum_{\alpha \neq \beta} n_{i \alpha \sigma} n_{i \beta \bar{\sigma}}\\
    & +\frac 1 2 (U^\prime - J_H) \sum_{i,\sigma} \sum_{\alpha \neq \beta} n_{i \alpha \sigma} n_{i \beta \sigma}\\
    & - J_H \sum_{i, \alpha \neq \beta} ( c^\dagger_{i \alpha \uparrow} c_{i \alpha \downarrow} c^\dagger_{i, \beta \downarrow} c_{i \beta \uparrow})\\&+J_H \sum_{i, \alpha \neq \beta} ( c^\dagger_{i \alpha \uparrow} c^\dagger_{i \alpha \downarrow} c_{i \beta \downarrow} c_{i \beta \uparrow})
 \label{Hint}
\end{align}
with Coulomb interaction $U$, $U^\prime$ and Hund's coupling $J_{H}$ connected
via $U^\prime = U-2J_{H}$. We use here also NN hopping $t\approx
0.2\;\textrm{eV}$ as unit of energy. Remaining parameters are $\Delta =
-0.5\;t$ (favoring $xy$ occupation), $U=14\;t$ and $J_{H} = 2\;t$. The
VCA was here based on a plaquette of size $2\times 2$ sites, we restrict ourselves to a plane modeling a surface of interaface.

\end{document}